\documentclass[reprint,aps,prl,twocolumn,amsmath,amssymb,amsfonts]{revtex4-2} %superscriptaddress,10pt,letterpaper
\usepackage[english]{babel}
\usepackage[utf8]{inputenc}
\usepackage{graphicx}
\usepackage{amsmath,amssymb,amsthm,mathtools}
\usepackage{txfonts}
\usepackage{xcolor}
\usepackage{bm}
\usepackage[normalem]{ulem}
\usepackage{textcomp}
\usepackage{soul}

\newcommand{\ket}[1]{\left| #1 \right \rangle}

\begin{document}

\title{Order-invariant two-photon quantum correlations in PT-symmetric interferometers}

\author{Tom~A.~W. Wolterink}
\email[]{tom.wolterink@uni-rostock.de}
\affiliation{Institute for Physics, University of Rostock, Albert-Einstein-Stra{\ss}e 23, 18059 Rostock, Germany}
\author{Matthias Heinrich}
\affiliation{Institute for Physics, University of Rostock, Albert-Einstein-Stra{\ss}e 23, 18059 Rostock, Germany}
\author{Stefan Scheel}
\affiliation{Institute for Physics, University of Rostock, Albert-Einstein-Stra{\ss}e 23, 18059 Rostock, Germany}
\author{Alexander Szameit}
\email[]{alexander.szameit@uni-rostock.de}
\affiliation{Institute for Physics, University of Rostock, Albert-Einstein-Stra{\ss}e 23, 18059 Rostock, Germany}

\begin{abstract}
	Multiphoton correlations in linear photonic quantum networks are governed by matrix permanents. Yet, surprisingly few systematic properties of these crucial algebraic objects are known, while their calculation is a computationally hard task. As such, predicting the overall multiphoton behavior of a network from its individual building blocks typically defies intuition. In this work we identify sequences of concatenated two-mode linear optical transformations whose two-photon behavior is invariant under reversal of the order. We experimentally verify this systematic behavior in parity-time-symmetric complex interferometer arrangements of varying composition. Our results underline new ways in which quantum correlations may be preserved in counterintuitive ways even in small-scale non-Hermitian networks.
\end{abstract}
\maketitle

\section{Introduction}\label{sec:introduction}
Quantum correlations in linear optical networks are a crucial resource for quantum information processing. In particular, the correlations between multiple photons are described by the permanent of the transmission matrix of the optical network \cite{Scheel2004,*Scheel2008}. Yet, despite their central role, very few systematic properties of permanents have been identified thus far \cite{Minc1978}. In stark contrast to the determinant of a matrix, the computational effort of calculating its permanent scales quite unfavorably with the dimensionality of the matrix itself \cite{Valiant1979}. At large scale, these twin limitations have sparked the field of boson sampling \cite{Aaronson2011,Broome2013,Spring2013,Tillmann2013,Crespi2013} through the connection to linear optical quantum computing. However, the lack of tangible permanent properties also affects the design of small-scale networks, as it hinders intuitive prediction of the behavior of composite systems from their constituent building blocks. This especially holds true for non-Hermitian systems that incorporate losses, which are known to drastically alter quantum correlations even in the simplest networks of just two modes, such as a lossy beam splitter \cite{Barnett1998,Jeffers2000,Uppu2016,Defienne2016,Wolterink2016,Vest2017}. 

Among the wide variety of non-Hermitian settings, systems obeying parity-time (PT) symmetry \cite{Bender1998} are of particular interest, since they can still possess entirely real-valued spectra despite violations of energy conservation.
PT-symmetric systems are described by Hamiltonians that are invariant under the combined operation of parity-inversion and time-reversal \cite{Bender2007}. Through tuning of a single physical parameter, they can undergo a symmetry-breaking phase transition at an exceptional point \cite{Heiss2004,Berry2004} where the eigenvalue spectrum becomes complex. 
Photonics provides an excellent platform to study PT symmetry \cite{ElGanainy2018,Miri2019,Ozdemir2019,Parto2021}, implementing non-Hermiticity through tailored gain and loss \cite{Ruschhaupt2005,ElGanainy2007,Makris2008,Musslimani2008}. Along these lines, light-based realizations have been instrumental for the observation of many features of PT symmetry in the classical domain \cite{Guo2009,Rueter2010}, with potential applications ranging from sensors with enhanced sensitivity \cite{Hodaei2017,Chen2017} to efficient lasers with robust single-mode characteristics \cite{Hodaei2014,Feng2014}. 
Notably, realizing PT-symmetric photonic systems in the quantum domain necessitates a shift to passive systems \cite{Ornigotti2014}, as gain-induced quantum noise would inevitably break PT symmetry \cite{Scheel2018} regardless of the symmetry of the underlying refractive index landscape. Such passive PT systems recently enabled the very first observations of PT-symmetric quantum interference \cite{Klauck2019} and the quantum simulation of coupled PT-symmetric Hamiltonians \cite{Maraviglia2022} on a photonic platform, opening up the exploration of quantum correlations in larger non-Hermitian networks. 

Here, we systematically identify types of sequences of two-mode systems that perform distinct linear optical transformations, whereas their permanents remain invariant under reversal of the entire sequence’s order. We experimentally verify this discovery by probing and comparing the two-photon correlations in PT-symmetric interferometers of wildly different composition, and demonstrate that quantum correlations in PT-symmetric photonic networks may be preserved in a counterintuitive manner. 

\section{Concatenated two-mode transformations}\label{sec:concatenatedtwomodetransformations}
Let us consider an arbitrary two-mode linear optical system whose action on a two-photon state is described by its $2\times2$ transmission matrix $M=\begin{psmallmatrix}m_{11} & m_{12}\\m_{21} & m_{22}\end{psmallmatrix}$. Its constituent elements $m_{ij}$ represent transmission and reflection between the two respective channels under the action of the system. When exciting this system with a $\ket{11}$ Fock state, the probability of observing coincidences between the outputs is given by $P_{11}= \left|\mathrm{perm}\,{M}\right|^2$ \cite{Scheel2004}. Notably, the same holds true for the row-and-column-reversed arrangement of $m_{ij}$ obtained by $XM^{\mathrm{T}}X=\begin{psmallmatrix}m_{22} & m_{12}\\m_{21} & m_{11}\end{psmallmatrix}$, where $X=\begin{psmallmatrix}0 & 1\\1 & 0\end{psmallmatrix}$ is the exchange matrix and $\vphantom{M}^{\mathrm{T}}$ denotes transposition.

\begin{figure}[ht!]
\includegraphics[scale=1]{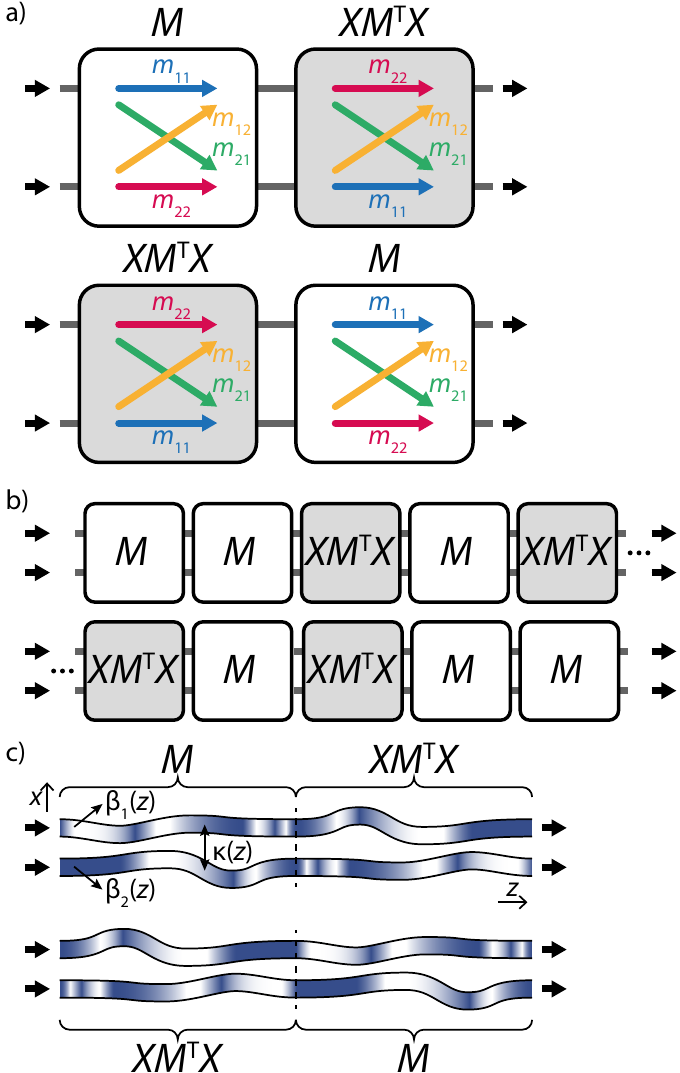}
\caption{Schematic representation of sequences of two-mode transformation $M$ and its row-and-column-reversed transpose $XM^{\mathrm{T}}X$ with their transmission/reflection coefficients $m_{ij}$. Exciting sequences of a) two or b) more copies of $M$ and $XM^{\mathrm{T}}X$, concatenated in either forward or backward order, with a $\ket{11}$ state gives rise to identical two-photon coincidences. c) Photonic implementation of the transformations $(XM^{\mathrm{T}}X)\,M$ and $M\,(XM^{\mathrm{T}}X)$  as two coupled waveguides with $z$-dependent propagation constants $\beta_1$, $\beta_2$ (shading) and coupling constant $\kappa$ (distance).}
\label{fig:figsketch}
\end{figure}

As the transmission matrices of these two systems do not commute, concatenating them can result in two different transformations, depending on the specific order, as illustrated in Fig.~\ref{fig:figsketch}(a)):
\begin{align}
\left(XM^{\mathrm{T}}X \right)M &= \begin{pmatrix}m_{11}m_{22}+m_{12}m_{21} & 2m_{12}m_{22}\\2m_{11}m_{21} & m_{11}m_{22}+m_{12}m_{21}\end{pmatrix} \label{eq:XMtXM} \\
M\left(XM^{\mathrm{T}}X\right) &= \begin{pmatrix}m_{11}m_{22}+m_{12}m_{21} & 2m_{11}m_{12}\\2m_{21}m_{22} & m_{11}m_{22}+m_{12}m_{21}\end{pmatrix} \label{eq:MXMtX}
\end{align}
Surprisingly, while these two transformations differ in their off-diagonal elements, their permanents, and with them the coincidence probability amplitudes for indistinguishable photons, are the same:
\begin{multline}
\mathrm{perm}\,{\left\lbrace\left(XM^{\mathrm{T}}X\right)M\right\rbrace}=\mathrm{perm}\,{\left\lbrace M\left(XM^{\mathrm{T}}X\right)\right\rbrace}= \\
{\left(m_{11}m_{22}+m_{11}m_{12}\right)}^2 + 4m_{11}m_{12}m_{21}m_{22}
\label{eq:perm1}
\end{multline}
In other words: while these two systems behave differently when probed with classical light or single photons, their two-photon characteristics are strictly identical. This fact is independent of the (in)distinguishability of the impinging photons, as also the probabilities for distinguishable photons are equal according to $\mathrm{perm}\,{\left|(XM^{\mathrm{T}}X)\,M\right|^2}=\mathrm{perm}\,{\left|M\,(XM^{\mathrm{T}}X)\right|^2}$, where the absolute square represents the Hadamard product of a matrix and its complex conjugate  $\left|A\right|^2 = A \circ A^{*}$.

%-------------------------
%-------------------------

Notably, this peculiar effect is not rooted in trivial properties of permanents such as their invariance to transposition or the permutation of rows or columns \cite {Minc1978}, but instead arises from the equal antidiagonal elements of $M$ and $XM^{\mathrm{T}}X$. From a physical point of view, it can be understood by tracing all possible exchange paths between input and output states in the two modes (see Fig.~\ref{fig:figsketch}(a)). To observe the state $\ket{11}$ at the output when $\ket{11}$ is injected at the input, both photons either need to remain in their modes, or whenever a photon would exchange modes, one needs to return. For systems with identical antidiagonals, this exchange always results in a probability of $m_{12}m_{21}$, regardless of where the flip occurs.
This property straightforwardly extends to longer sequences, as for any sequence of complex-valued $2\times2$ matrices with equal antidiagonals, multiplication to either the left or the right results in matrices with identical diagonals and permanents. 
Thus, the permanent of any arbitrary sequence of transformations $M$ and $XM^{\mathrm{T}}X$, and therefore its two-photon behavior, is strictly invariant under reversal of the entire sequence (Fig.~\ref{fig:figsketch}(b)). Note that this invariance does not directly translate to systems with a larger number of modes or photons. Instead, e.g. in a three-mode system of concatenated transformations $M$ and $N$, where $N$ has the same entries as $M$, the only matrices whose permanents or subpermanents do not depend on the order are related through $N=PMP$, with a permutation matrix $P$. In those cases, the (sub)permanents of $N\,M$ and $M\,N$, and thus their multiphoton behavior, are trivially equal. Whether order-invariant photon correlations systematically occur in larger systems or may be enforced by certain symmetries remains an open question.

When we interpret the transformation $M$ as the result of the evolution of a time-dependent Hamiltonian $H(z)$,
\begin{equation}
M=\mathcal{T}\exp{\left(-i \int_{0}^{l} H(z)\,dz\right)}
\label{eq:timeevolution}
\end{equation}
its row-and-column-reversed counterpart $XM^{\mathrm{T}}X$ corresponds to evolving $H(z)$ in reverse order while also swapping parity. Note that this is distinct from a conventional time reversal, as it involves no complex conjugation.
The concatenations $(XM^{\mathrm{T}}X)\,M$ and $M\,(XM^{\mathrm{T}}X)$ then correspond to the evolution of a Hamiltonian which reverses its order and swaps parity midway, having point symmetry in the $x$-$z$ plane, and its permanent is invariant to reversing the order of the first and second halves.
The invariance of the permanent holds for any complex-valued $2\times2$ transmission matrix $M$ and thus naturally includes non-Hermitian systems. For the subset of unitary systems, the transformations $(XM^{\mathrm{T}}X)\,M$ and $M\,(XM^{\mathrm{T}}X)$ only differ in external phases that photon-counting statistics are fundamentally agnostic towards, so that losses are in fact essential to distinguish non-trivially invariant two-photon correlations. 

\section{PT-symmetric interferometers}\label{sec:PTsymmetricinterferometers}
In a photonic context, such Hamiltonians can be readily mapped onto a system of two interacting waveguides with complex on-site potentials, all of which may be varying along the propagation coordinate $z$ in ways that obey the symmetry condition as sketched in Fig.~\ref{fig:figsketch}(c). In the following, we will therefore turn to customized waveguide circuits to probe this invariance in PT-symmetric interferometer arrangements. Their fundamental building block, a passive PT coupler, consists of two waveguides interacting with a coupling constant $\kappa$ over a certain length $l$, while one of the guides is subject to losses at a rate of $\gamma$. Evolution in the system is governed by the effective Hamiltonian:
\begin{equation}
H_{\mathrm{eff}}=\begin{pmatrix}-i\gamma && \kappa \\ \kappa && 0\end{pmatrix}
\label{eq:matH}
\end{equation}
For $\gamma/\kappa<2$, PT symmetry is unbroken and the Hamiltonian has real eigenvalues when viewed in the co-damped reference frame \cite{Ornigotti2014}. The full quantum-mechanical evolution of the non-Hermitian system may be described using a noise-operator approach \cite{Matloob1995}, using a quantum master equation in Lindblad form \cite{Manzano2020,Teuber2020}, or by unitary dilation \cite{Halmos1950}. When one probes the two-photon behavior by post-selecting the cases where neither of the photons is lost, the probability amplitudes directly follow from the classical propagator $\tilde{U}=\exp\left(-iH_{\mathrm{eff}}l\right)$, which now describes a lossy (i.e. non-unitary) transformation.

%-------------------------
\begin{figure}[ht!]
\includegraphics[scale=1]{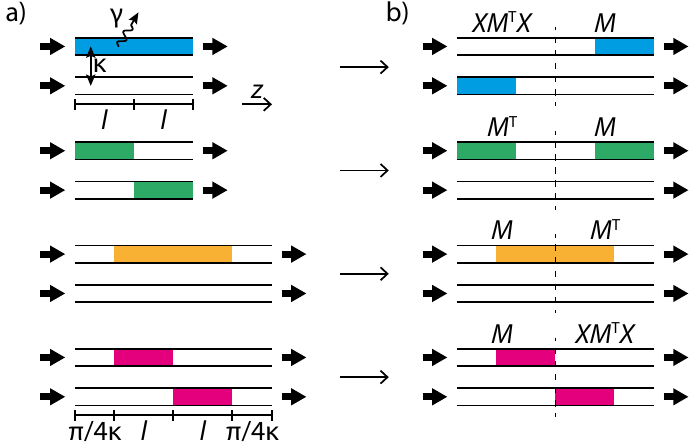}
\caption{Schematic drawing of a) concatenated PT-symmetric directional couplers (length $l$) and conventional 50:50 directional couplers ($\gamma=0$, length $\pi/4\kappa$) and b) their equivalent descriptions in terms of the transformation $M$, consisting of a 50:50 directional coupler followed by a PT coupler. Color of the lossy section is used to distinguish the system geometry: $M\,(XM^{\mathrm{T}}X)$ (cyan), $M\,M^{\mathrm{T}}$ (green), $M^{\mathrm{T}}\,M$ (yellow), $(XM^{\mathrm{T}}X)\,M$ (magenta).}
\label{fig:figsetup}
\end{figure}
%-------------------------

\begin{figure}[ht!]
\includegraphics[scale=1]{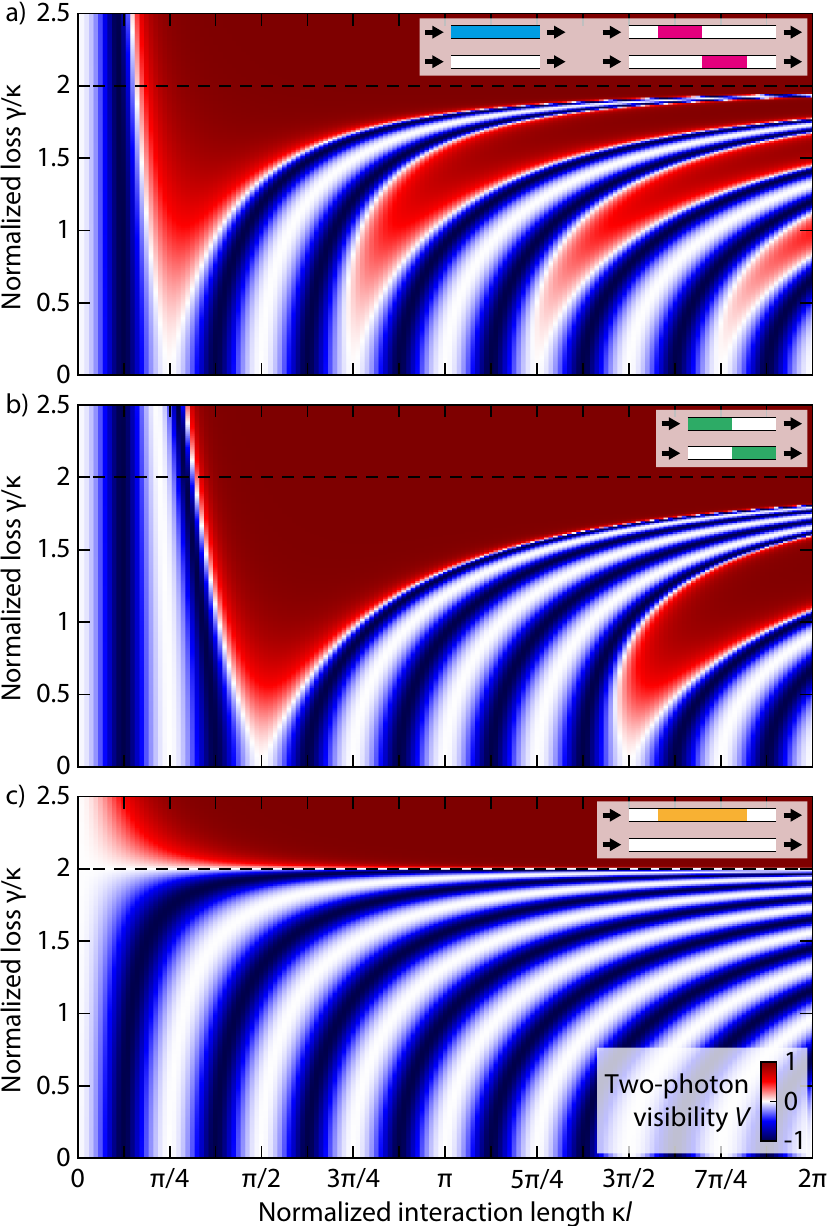}
\caption{Calculated two-photon visibility as a function of normalized length $\kappa l$ and loss coefficients $\gamma/\kappa$ in the non-Hermitian systems with geometries of a) $M\,X(M^{\mathrm{T}}X)$ and $(XM^{\mathrm{T}}X)\,M$, b) $M\,M^{\mathrm{T}}$, c) $M^{\mathrm{T}}\,M$. Spontaneous PT-symmetry breaking occurs at $\gamma/\kappa=2$. Negative visibility corresponds to relative bunching of photons, with complete HOM interference at $V=-1$, and positive visibility to antibunching behavior.}
\label{fig:figtheory}
\end{figure}

We now construct interferometers of two concatenated PT couplers that are either aligned or inverted with respect to their loss distribution (Fig.~\ref{fig:figsetup}(a)). The aligned sequence (cyan) yields a single PT coupler of twice the length, i.e. length $2l$. The inverted arrangement with opposite loss profile (green) nevertheless remains PT symmetric at each point along $z$. Additionally, we consider both of these systems in a rotated basis \cite{Ehrhardt2022} realized by placing them between a pair of Hermitian directional 50:50 couplers (yellow and magenta, respectively). As it turns out, each of these four systems can be equivalently described in terms of a transformation $M=\tilde{U}R$ that consists of a 50:50 directional coupler $R=\frac{1}{\sqrt{2}}\begin{psmallmatrix}1 && -i \\ -i && 1\end{psmallmatrix}$ followed by a PT coupler $\tilde{U}$. Since $R^2=-iX$, two successive 50:50 couplers bring about a full population transfer, effectively swapping parity while adding an otherwise inconsequential global phase. Figure~\ref{fig:figsetup}(b) displays the four configurations of the systems thus expanded to the same overall interaction length: $M\,(XM^{\mathrm{T}}X)$, $M\,M^{\mathrm{T}}$, $M^{\mathrm{T}}\,M$, and $(XM^{\mathrm{T}}X)\,M$, respectively. These representations reveal that sandwiching the aligned or inverted PT coupler sequences in between two directional couplers indeed establishes the aforementioned permanent-preserving symmetry condition between those otherwise very different arrangements and, by extension, imbues them with identical two-photon behavior. This can readily be verified by calculating the two-photon visibility of quantum interference, defined as $V = P_{\mathrm{indist}}/ P_{\mathrm{dist}}-1$ with $P_{\mathrm{indist}}$ and $P_{\mathrm{dist}}$ the coincidence probability for indistinguishable and distinguishable photons, respectively, as a function of the normalized loss $\gamma/\kappa$ and interaction length $\kappa l$, as shown in Fig.~\ref{fig:figtheory}.

%-------------------------
%-------------------------

Figure~\ref{fig:figtheory}(a) shows the two-photon visibility for first configuration of aligned PT couplers, $M\,(XM^{\mathrm{T}}X)$, that is equivalent to a single PT coupler of length $2l$. In the lossless case ($\gamma/\kappa=0$), the visibility periodically oscillates between $-1$ and $0$ as a function of length as the effective splitting ratio of the coupler varies, with the first Hong-Ou-Mandel (HOM) dip \cite{Hong1987} occurring at $\kappa l=\pi/8$. Introducing loss into the system ($\gamma/\kappa<2$) systematically alters these dynamics by simultaneously slowing down the overall oscillation period and shifting the first minimum towards shorter interaction lengths  \cite{Klauck2019}. Furthermore, while the minima remain at -1, the visibility now oscillates between negative (bunching) and positive (anti-bunching) as a function of length: Around the odd-numbered maxima of $V$, associated with a 0:100 coupler in the Hermitian case, the visibility now turns positive and thus crosses zero twice, indicating that two-photon interference actually increases the survival probability of the photon pair. Whenever one of the transmission/reflection coefficients of the system crosses zero and changes sign, the visibility reaches zero as well, because the quantum interference vanishes, and likewise swaps sign afterwards. At even maxima of $V$, two coefficients simultaneously cross zero, such that the sign of the visibility remains unaffected.
In contrast, above the PT-symmetry threshold ($\gamma/\kappa>2$), the system's oscillation period turns imaginary, erasing all but the first HOM dip as $V$ monotonically approaches $+1$ for all larger interaction lengths $\kappa l$. Likewise, the $(XM^{\mathrm{T}}X)\,M$ configuration of inverted PT couplers between two directional couplers obeys the permanent-preserving symmetry and therefore displays an identical two-photon behavior for any choice of effective gain or interaction distance.

Note that antibunching behavior ($V>0$) is precluded in the lossless system and can only be achieved in the non-Hermitian context. Even though PT-symmetric directional couplers and, more generally, lossy directional couplers, are both non-Hermitian systems that can exhibit superficially similar oscillations in the two-photon visibility, two distinct underlying mechanisms are at work. In lossy directional couplers, the sign and magnitude of $V$ depend on both the amplitudes of the transmission/reflection coefficients as well as an internal phase \cite{Barnett1998,Jeffers2000,Uppu2016}. Zero visibility, for example, can therefore arise from the absence of interference, when one of the coefficients reaches zero, or dynamically occur through interference for specific values of the internal phase. In contrast, PT symmetry restricts the value of the internal phase, such that the magnitude of $V$ solely depends on the amplitudes of the coefficients: zero visibility thus always corresponds to the complete suppression of interference.

Figure~\ref{fig:figtheory}(b) illustrates the two-photon visibility in inverted PT coupler sequences ($M\,M^{\mathrm{T}}$). Qualitatively, the influence of loss on the system is similar to the aligned case. However, the dependence on the section length is different, as the visibility now turns positive every four maxima, starting from the second one. Upon closer inspection, the locations of the even maxima and the sign of the visibility correspond to those of a single PT coupler (cf. Fig.~\ref{fig:figtheory}(a), but scaled by a factor of 2 in horizontal direction for a single coupler of length $l$.) This can be understood by considering that zeros in the transmission/reflection coefficients of a single coupler directly map into zero coefficients, and thus vanishing $V$ in the inverted sequence.
Moreover, additional (odd) maxima of zero visibility appear at exactly those lengths where full HOM bunching (minimum visibility) occurs in a single coupler, as the second coupler reverses the transformation of the first, leading to pairs of zero coefficients and vanishing interference.

Finally, two aligned PT couplers, or, equivalently, a single one of twice the length, sandwiched between two 50:50 directional couplers, $M^{\mathrm{T}}\,M$ exhibits an entirely different behavior (Fig.~\ref{fig:figtheory}(c)). In the unbroken phase ($\gamma/\kappa<2$), the visibility stays strictly negative despite oscillating at similarly increasing periods, remains identically zero regardless of the interaction length at the PT-breaking threshold ($\gamma/\kappa=0$), and turns globally positive in the broken phase ($\gamma/\kappa>2$). As such, this type of arrangement allows for the PT-broken phase to be unambiguously identified directly from its quantum correlations \cite{Longhi2020}.

\section{Experimental observations}\label{sec:experiments}
\begin{figure}[ht!]
\includegraphics[scale=1]{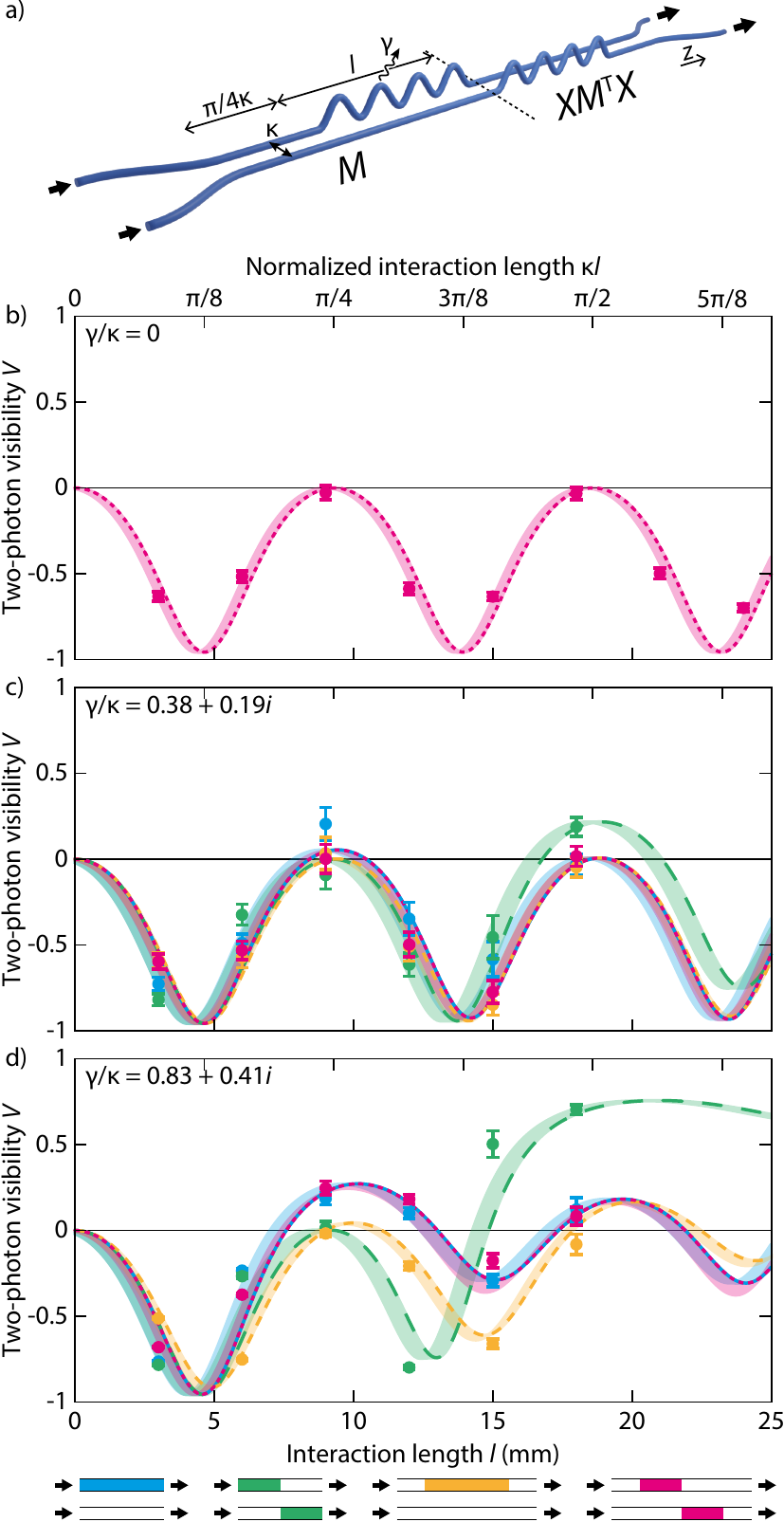}
\caption{a) Experimental implementation of $XM^{\mathrm{T}}X\,M$, implementing the desired loss through modulation of waveguide trajectory. b-d) Measured (points) and calculated (curves) two-photon visibility as a function of length $l$ in the non-Hermitian systems, for increasing loss coefficients of b) $\gamma/\kappa = 0$ (lossless), c) $\gamma/\kappa = 0.38+0.19i$, d) $\gamma/\kappa = 0.83+0.41i$. Color indicates the system geometry: $M\,(XM^{\mathrm{T}}X)$ (cyan), $M\,M^{\mathrm{T}}$ (green), $M^{\mathrm{T}}\,M$ (yellow), $(XM^{\mathrm{T}}X)\,M$ (magenta). Error bars are based on the square root of the number of observed coincidences. The shaded areas indicate the effect of additional coupling in the fanning sections.}
\label{fig:figexp}
\end{figure}

To experimentally test the predictions of our model, we fabricate non-Hermitian waveguide circuits via the femtosecond-laser direct writing technique \cite{Szameit2010}. The individual channels are inscribed by focusing 270~fs laser pulses from an ultrafast fiber laser amplifier (Coherent Monaco,  wavelength 512~nm) at a repetition rate of 333~kHz and an average power of 70~mW through a 50$\times$ microscope objective (NA=0.6) into fused silica (Corning 7980). The waveguide trajectories are defined by the motion of a high-precision translation stage (Aerotech ALS180) at a speed of 100~mm/min. At the design wavelength of 814~nm, the propagation losses of these waveguides are below 0.12~dB~cm$^{\mathrm{-1}}$. In the interaction regions of our couplers, the waveguides are separated by 20~{\textmu}m, corresponding to a coupling coefficient of $\kappa=0.85~\mathrm{cm^{-1}}$. The desired losses are introduced into the waveguides by rapidly undulating their out-of-plane positions following a cosine trajectory \cite{Eichelkraut2014} with a 0.15~cm period and amplitudes of 1.0~{\textmu}m and 1.5~{\textmu}m, resulting in excess loss coefficients of $\gamma=0.32~\mathrm{cm^{-1}}$ and $0.70~\mathrm{cm^{-1}}$, respectively. Figure~\ref{fig:figexp}(a) schematically illustrates the waveguide geometry used to implement the $(XM^{\mathrm{T}}X)\,M$ configuration. We then probe the two-photon dynamics in these systems by injecting horizontally polarized wavelength-degenerate photon pairs at 814~nm generated by type-I spontaneous parametric down-conversion from a continuous-wave pump at 407~nm in a bismuth borate crystal and registering coincidence counts between the channels. The degree of indistinguishability of the photons was characterized through observing HOM interference, resulting in a visibility of 96\%.

As reference for the non-Hermitian arrangements, Fig.~\ref{fig:figexp}(b) shows the experimentally observed visibility of two-photon quantum interference as a function of the length in a conventional Hermitian coupler ($\gamma/\kappa = 0$): a clear sequence of identical HOM dips unfolds as the effective splitting ratio of the coupler varies with its length. Note that this conventional coupler is the limiting case of all four complex configurations arrangements for vanishing losses. As soon as losses are introduced, differences in the visibility dynamics become apparent. Figure~\ref{fig:figexp}(c) shows the observed behavior for a loss coefficient of $\gamma/\kappa = 0.38+0.19i$. While all four PT-symmetric configurations retain certain similarities at such low loss levels and $V$ stays negative for $M^{\mathrm{T}}\,M$ (yellow), $M\,(XM^{\mathrm{T}}X)$ (cyan) and $(XM^{\mathrm{T}}X)\,M$ (magenta) both 
begin to display positive visibility around the first maximum, whereas $M\,M^{\mathrm{T}}$ (green) turns positive at the second one.

As shown in Fig.~\ref{fig:figexp}(d), the qualitative differences in the behavior of the four configurations become more prominent for increased losses ($\gamma/\kappa = 0.83+0.41i$), as  $M^{\mathrm{T}}\,M$ (yellow) and $M\,M^{\mathrm{T}}$ (green) systematically diverge from $M\,(XM^{\mathrm{T}}X)$ (cyan) and $(XM^{\mathrm{T}}X)\,M$ (magenta), whose two-photon visibility dynamics are both governed by the same matrix permanent. The measurements (data points) are in good agreement with the predicted behavior as well as calculations based on system parameters retrieved in a classical calibration (curves), with minor deviations attributable to additional coupling in the fanning sections (shading) and inadvertent detunings of the propagation constant incurred by the undulating loss regions. The latter results in complex-valued loss coefficients $\gamma/\kappa$ which gradually reduce the interference contrast such that $V$ tends to zero at longer propagation distances. Notably, even under these imperfect conditions, the two-photon visibilities of  $(XM^{\mathrm{T}}X)\,M$ and $M\,(XM^{\mathrm{T}}X)$ are kept in lock-step by the permanent-preserving symmetry existing between them.

%-------------------------
%-------------------------

\section{Conclusion}\label{sec:conclusion}
In summary, we have identified a new type of symmetry transformation that preserves the two-photon interference properties in  sequences of non-Hermitian two-mode systems. From an algebraic point of view, this is a consequence of a property of matrix permanents, which remain invariant when transforming complex sequences of $2\times2$ matrices in line with this type of symmetry transformation. We have experimentally verified these findings in PT-symmetric interferometers of varying composition by demonstrating that two-photon correlations are indeed preserved by this symmetry. Whether networks with a larger number of modes may support similar order-invariant correlations remains an open question. Nevertheless, our results emphasize that even in deceptively simple two-mode systems, non-Hermitian quantum correlations may be governed by highly non-intuitive mechanisms, and that judiciously placed losses are in fact essential to distinguish non-trivially invariant two-photon correlations. The PT-symmetric interferometers investigated here pave the way for the incorporation of non-Hermitian building blocks into larger quantum photonic networks. Along these lines, we hope that our work will inspire a new approach to the design of linear optical networks that harness non-Hermiticity for advanced quantum information processing and sensing applications.

\begin{acknowledgments}
We thank C. Otto for preparing the high-quality fused-silica samples used in this work. This work was funded by Deutsche Forschungsgemeinschaft via SFB 1477 ``Light-Matter Interactions at Interfaces'', project no. 441234705. T.A.W.W. is supported by a European Commission Marie Skłodowska-Curie Actions Individual Fellowship ``Quantum correlations in PT-symmetric photonic integrated circuits'', project no. 895254.
\end{acknowledgments}

\bibliography{PTsp_references_20230202.bib}

\end{document}